\documentclass[final,3p,times,twocolumn]{elsarticle}

\usepackage{graphics}
\usepackage{graphicx}
\usepackage{amssymb}

\journal{Journal of Theoretical Biology}

\begin{document}

\begin{frontmatter}



\title{A Model of Sequential Branching in Hierarchical Cell Fate Determination}



\author{David V. Foster\corref{cor1}\fnref{ibi}}
\ead{dfoster@ucalgary.ca}
\cortext[cor1]{To whom correspondence should be addressed}
\author{Jacob G. Foster\fnref{comp}}
\author{Sui Huang\fnref{ibi}}
\author{Stuart A. Kauffman\fnref{ibi}}
\fntext[ibi]{Institute for Biocomplexity and Informatics, University of Calgary, Calgary, Alberta T2N 1N4}
\fntext[comp]{Complexity Science Group, Department of Physics and Astronomy, University of Calgary, Calgary, Alberta T2N 1N4}


\begin{abstract}
Multi-potent stem or progenitor cells undergo a sequential series of binary fate decisions, which ultimately generate the diversity of differentiated cells. Efforts to understand cell fate control have focused on simple gene regulatory circuits that predict the presence of multiple stable states, bifurcations and switch-like transitions.  However, existing gene network models do not explain more complex properties of cell fate dynamics such as the hierarchical branching of developmental paths. Here, we construct a generic minimal model of the genetic regulatory network controlling cell fate determination, which exhibits five elementary characteristics of cell differentiation: stability, directionality, branching, exclusivity, and promiscuous expression. We argue that a modular architecture comprising repeated network elements reproduces these features of differentiation by sequentially repressing selected modules and hence restricting the dynamics to lower dimensional subspaces of the high-dimensional state space.  We implement our model both with ordinary differential equations (ODEs), to explore the role of bifurcations in producing the one-way character of differentiation, and with stochastic differential equations (SDEs), to demonstrate the effect of noise on the system. We further argue that binary cell fate decisions are prevalent in cell differentiation due to general features of the underlying dynamical system. This minimal model makes testable predictions about the structural basis for directional, discrete and diversifying cell phenotype development and thus can guide the evaluation of real gene regulatory networks that govern differentiation.
\end{abstract}

\begin{keyword}
cell differentiation \sep genetic regulatory network \sep dynamics \sep modularity


\end{keyword}

\end{frontmatter}


\section{Introduction}
The genetic regulatory network (GRN) plays an essential role in controlling cell differentiation and maintaining the stability of cell types. Each cell type has its own stable gene expression profile (the set of genes in its genome which are actively expressed) \cite{Kauffman_1969}. Thus, understanding cell differentiation requires a model of the behavior of the many interacting genes that establish the expression profile associated with each cell type.

Gene regulatory interactions are often described as a network, with each node representing a gene and links between two genes indicating a regulatory interaction. The large-scale characteristics of genetic regulatory networks have been studied in some detail \cite{Alon_2009, Davidson_2002}, but only in terms of network structure: the degree sequence \cite{Lee_2002}, the prevalence of network motifs \cite{Milo_2002, Shen-Orr_2002}, and so forth. In terms of GRN dynamics, most efforts have been ``bottom-up" --- devoted to simulating specific regulatory interactions or analyzing models of small genetic circuits \cite{Ackers_1982, Chickarmane_2006} and therefore focused on specific subsections of the network. Our goal is to work in a ``top-down" fashion to construct a minimal dynamical model which captures many important features of cell differentiation. Most significantly, we capture the sequential branching differentiation of cells from the multipotent progenitor cell to more lineage-restricted cells and finally to terminal cells.

To construct our model, we first note five elementary characteristics of cell differentiation which we wish to capture: stability, directionality, branching, exclusivity, and promiscuous expression. Of course, this is not meant to be an all-encompassing or fully detailed description of cell differentiation. However, these simplified characteristics prove rich enough to create an interesting model. We review here our definition of each characteristic.

\textit{Stability} : Cell types should be stable states of the dynamics. A cell which is in specific cell type should remain in that cell type for a ``long time'' (relative to the time scale of gene activation, transcription and translation), unless perturbed by an external signal. 

\textit{Directionality}: Progenitor cells should reliably differentiate into a variety of more specialized cell types \cite{Huang_2005} and should only very rarely regain pluripotency by de-differentiating \cite{Kal_2004}.

\textit{Branching}: Differentiation should occur through discrete lineage specifications, in which a multipotent cell makes a binary fate decision and commits to one of two different lineages. 

\textit{Exclusivity}: Once cells in our model have committed to a specific lineage, they should lose the ability to produce cell types from the other lineage. Each differentiation event both selects a specific lineage and forecloses the possibility of the other.

\textit{Promiscuous expression}: Multipotent progenitor cells should simultaneously express genes specific to all of their prospective lineages. For example, a hematopoietic stem cell expresses genes found in all of the lineages into which it can ultimately differentiate \cite{Cross_1997}.

\section{Model specification}
\begin{figure}
\includegraphics[scale=.21]{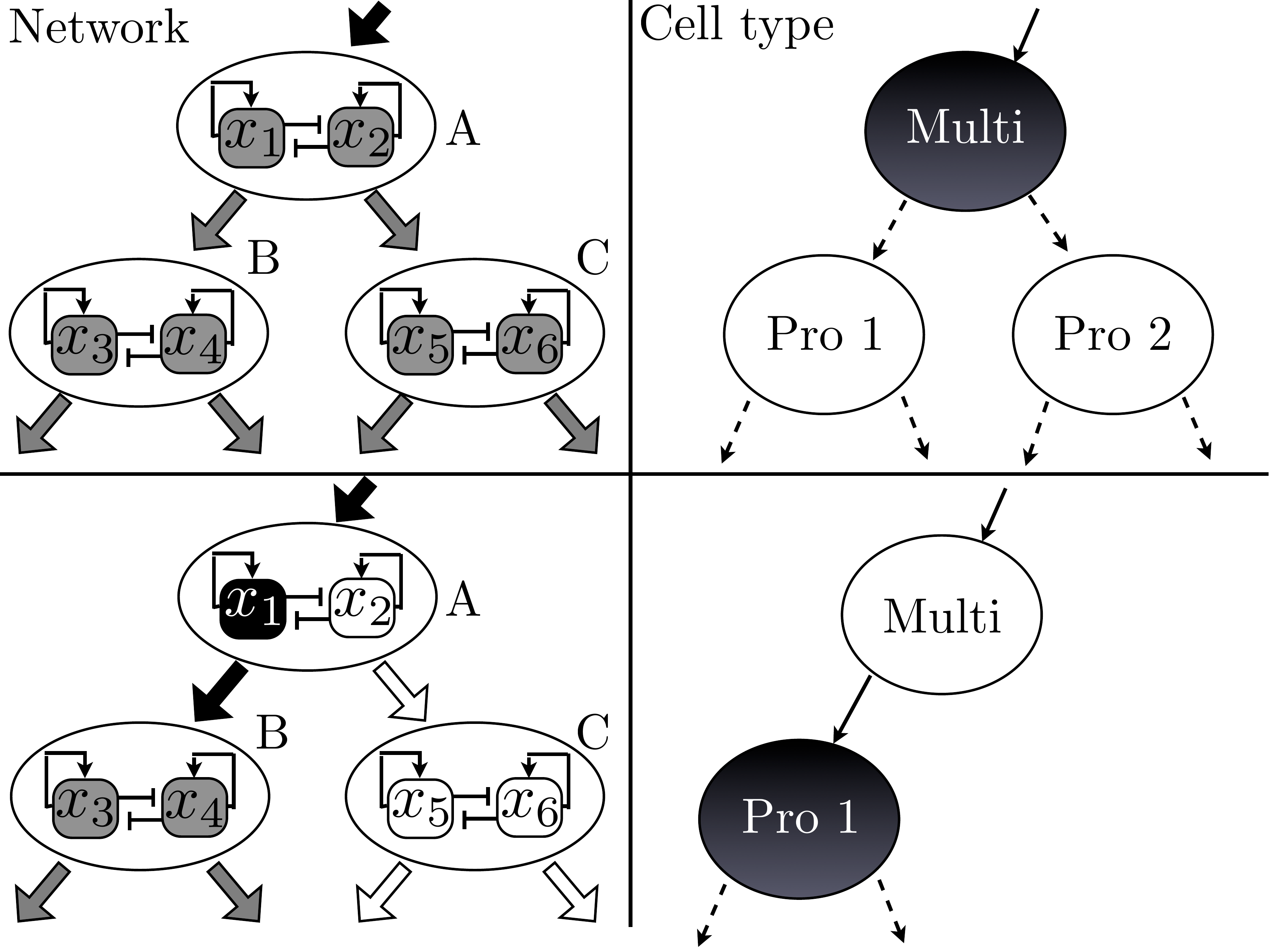}
\caption{\label{Biftree} Transition of a module in our network from the multipotent cell type to one of the progenitor cell types. Modules consist of two genes, each of which activates its own transcription and inhibits the transcription of the other gene. In the top panels, both genes in module A are active at a moderate level, represented by the gray coloration. This intermediate steady state is poised between the two subordinate lineages. Modules B and C are also both expressing their genes at a moderate level, as a result of the influence of module A. In the lower panels, the cell has differentiated from the multipotent cell type to progenitor type 1. Module A has switched from the intermediate stable state to the stable state in which gene $x_1$ is highly active (as represented by the black coloration), whereas gene $x_2$ has become inactive (as represented by the white coloration). As a result, module C has become completely inactive due to the signals from module A, whereas module B remains in the intermediate steady state, and is thus prepared to make further cell fate decisions.}
\end{figure}

To create a model which exhibits these characteristics, we take our inspiration from the reported network properties of real GRNs. Biological networks exhibit modularity, i.e. they can be divided into largely independent subnetworks, called modules \cite{Hartwell_1999, Ihmels_2002, Ravasz_2002}.  The nodes in a given module interact strongly with other nodes in the module, but only weakly with nodes in other modules. In our model, the network is composed of many modules, each responsible for a specific cell fate decision. These modules influence each other in a hierarchically organized fashion, as shown in Fig. \ref{Biftree}, which ensures that differentiation events occur in a branching, sequential fashion. The dynamical state associated with each cell type is a result of the dynamics of the associated module. Arrows pointing from one module to another indicate influence of the former on the latter. Influence between modules acts as follows: a parameter in the regulated module depends on the expression level of a gene in the regulating module. In the top pairs of panels the system is in the ``Multi" (multipotent) state, and in the bottom pair, the system is in one of the two available progenitor cell types. On the dynamical side, the module A undergoes a change in its state (the specifics of which we will discuss shortly). In making the change, it exerts an influence on the two modules downstream, in this case turning off module C and allowing module B to enter a state which corresponds to ``Pro 1" (Progenitor 1). Notice that like module A, modules B and C are also linked to down-stream modules on which they exert influence, and that module A is itself influenced by a module ``higher up" on the differentiation cascade. This allows the process to proceed in series through as many levels as are required. Our model suggests an important role for modularity: the topology of the regulatory network sequentially restricts the trajectory of the dynamical system to ever smaller subspaces of the state space. These restrictions ensure that a given cell type can only take the appropriate pathways of subsequent differentiation. The dynamics of each module of the network imposes the branching structure on cell differentiations. 

\section{Dynamics of a single module}
Previous work has studied the behavior of GRNs using a variety of different dynamical system models: Boolean networks \cite{Espinosa-Soto_2004, Kauffman_2003, Kauffman_2004}, differential equations \cite{Chickarmane_2006, Gardner_2000}, and chemical master equations \cite{Danos_2008, Ribeiro_2006}. Although each model has its merits, for our current purposes differential equations provide the best combination of rich behavior and analytical tractability. 

\subsection{Deterministic model}
To model a set of $N$ interacting genes with differential equations, we associate the expression level (number of proteins produced by the gene which are present in the cell) of each gene with a variable $x_i$. The state of the system is given by the vector ${\bf x} = [x_1 \ldots x_N]$. If we plot the expression level of each gene as one axis in a Cartesian coordinate system, the state space of the dynamics---the space of all possible states ${\bf x}$---is $ \{  \mathbb{R}^+ \}^N$, where the restriction to the non-negative reals $\mathbb{R}^+$ follows from the fact that negative gene expression levels are unphysical. Since genes influence the transcription rates of other genes, the rate of transcription (or equally the instantaneous rate of change of the expression level) of each gene, $\dot{x_i}$,  is given by:
\begin{equation}
  \dot{x}_i = F_{i}({\bf x})
   \label{genODE}
\end{equation}
For a given state, the instantaneous rate of change of the expression level of each gene defines a vector ${\bf \dot{x}} = [\dot{x}_1 \ldots \dot{x}_N]$, which points in the direction of the future time evolution of the state under the dynamics.  These vectors trace out the trajectory of the system, i.e. the time series of states the system will visit.  
 
A fixed point is any point in state space in which the state of the system is ``trapped" under the deterministic dynamics (ignoring noise). Formally, this occurs for any point ${\bf x}^\star$ such that $\dot{x}_i^\star = F_{i}({\bf x}^\star) = 0$ for all $i$. The fixed point is stable to small perturbations if the eigenvalues of the Jacobian at ${\bf x}^\star$ are all negative. In our model, each cell type is represented by a stable fixed point of the dynamics. Our goal is to construct the simplest plausible set of genetic regulatory interactions for which the state of the system moves, either by random fluctuations or external signals, through a sequence of branching, exclusive steps from the fixed point which represents a multipotent cell type, to states which represent a series of intermediary cell types, and finally to a terminal cell type. 

Huang \textit{et al.} \cite{Huang_2007} provide experimental evidence that the genetic circuit controlling hematopoietic (blood cell) stem cells consists of a pair of mutually repressing genes, each of which also acts as a self-activator. Such a feedback loop is reported to be an over-represented motif in real genetic networks \cite{Lee_2002}. We have chosen the Huang switch for our cell-fate controlling module, as shown in Fig. \ref{Biftree}. The differential equations describing the Huang switch are:
\begin{equation}
\begin{array}{l}
\displaystyle \dot{x_1}  =  \frac{a_1 x_1^n}{x_1^n + \theta_{1}^n} + \frac{b_1 \psi_1^n}{x_2^n + \psi_{1}^n} - k_1 x_1\\
\\
\displaystyle \dot{x_2}  =  \frac{a_2 x_2^n}{x_2^n + \theta_{2}^n} + \frac{b_2 \psi_2^n}{x_1^n + \psi_{2}^n} - k_2 x_2\\
\end{array} 
\label{diffeq}
\end{equation}
These equations model the dynamics of self-activation and mutual inhibition with Hill functions. The parameters $a_1, a_2$ give the maximum transcription rate due to self-activation, and $\theta_1, \theta_2$ control the concentration of protein $x_1$ or $x_2$ at which the inflection point in the Hill function occurs. Parameters $b_1, b_2$ give the basal rate of transcription, which is suppressed by a downward sloping Hill function with $\psi_1, \psi_2$ controlling the concentration of protein at which its inflection point occurs. The cooperativity of the interactions is given by $n$, which governs the steepness of the Hill function. The rate of protein degradation is controlled by $k_1, k_2$. 

To simplify our analysis we impose certain symmetries on the equations, and select specific values for certain parameters to produce the desired behavior.  We make the maximum contributions of the basal expression rate and the self-activation the same ($a_i = b_i$) and thus have a single parameter which controls the overall maximum transcription rate of the gene. For simplicity, we also set the inflection points of the two Hill functions (for the self-activation and inhibition) to the same value ($\theta_i = \psi_i = \theta$). Finally, we choose to make all parameter values $a$, $\theta$, and $k$ identical between the two equations (so $k_1 = k_2$, etc.), thus making the equations symmetric between $x_1$ and $x_2$. We can write the new set of parameter values as: 
\begin{equation}
\begin{array}{l}
\displaystyle a_1, a_2, b_1, b_2  = a \\
\displaystyle \theta_1, \theta_2, \psi_1, \psi_2 =  \theta = 1 \\
\displaystyle n = 4\\
\displaystyle k_1, k_2 = k \\
\end{array} 
\end{equation}
where we have chosen to set the inflection point $\theta = 1$ and the cooperativity $n = 4$. These parameter choices are not unique, and do not require careful tuning: as long as $n \geq 4$ a wide range of values of $\theta$, $a$, and $k$ produce the desired tri-stable behavior. This leaves us with the new simplified equations:
\begin{equation}
\begin{array}{l}
\displaystyle \dot{x_1}  = a \left( \frac{x_1^4}{x_1^4 + 1} + \frac{1}{x_2^4 + 1} \right) - k x_1\\
\\
\displaystyle \dot{x_2}  = a \left( \frac{x_2^4}{x_2^4 + 1} + \frac{1}{x_1^4 + 1} \right) - k x_2\\
\end{array} 
\label{simplediffeq}
\end{equation}
These equations have only two free parameters: $a$ and $k$. These are the parameters which are influenced by other modules, following the schematic shown in Fig \ref{Biftree}, and as detailed below. In order to better understand the effect of these parameters, we explore the bifurcation that occurs as they are changed. A bifurcation is a dynamical event in which a small change of a parameter leads to a qualitative change in the dynamics of the system. Bifurcations can destabilize fixed points, or cause points which were previously unstable to become stable. Thus bifurcations that change the stability of a fixed point are the dynamical system analogues of cell fate decisions in differentiating cells, where we represent cell types as fixed points for simplicity. For example, a parameter change may destabilize the fixed point representing a multipotent cell type, thus driving the state to a different fixed point, which represents the new progenitor cell type. Note that while a fixed point may be stable under the deterministic dynamics of the system, in the presence of noise even a stable fixed point may not trap the system indefinitely; see the next subsection.

\begin{figure}
\includegraphics[scale=.21]{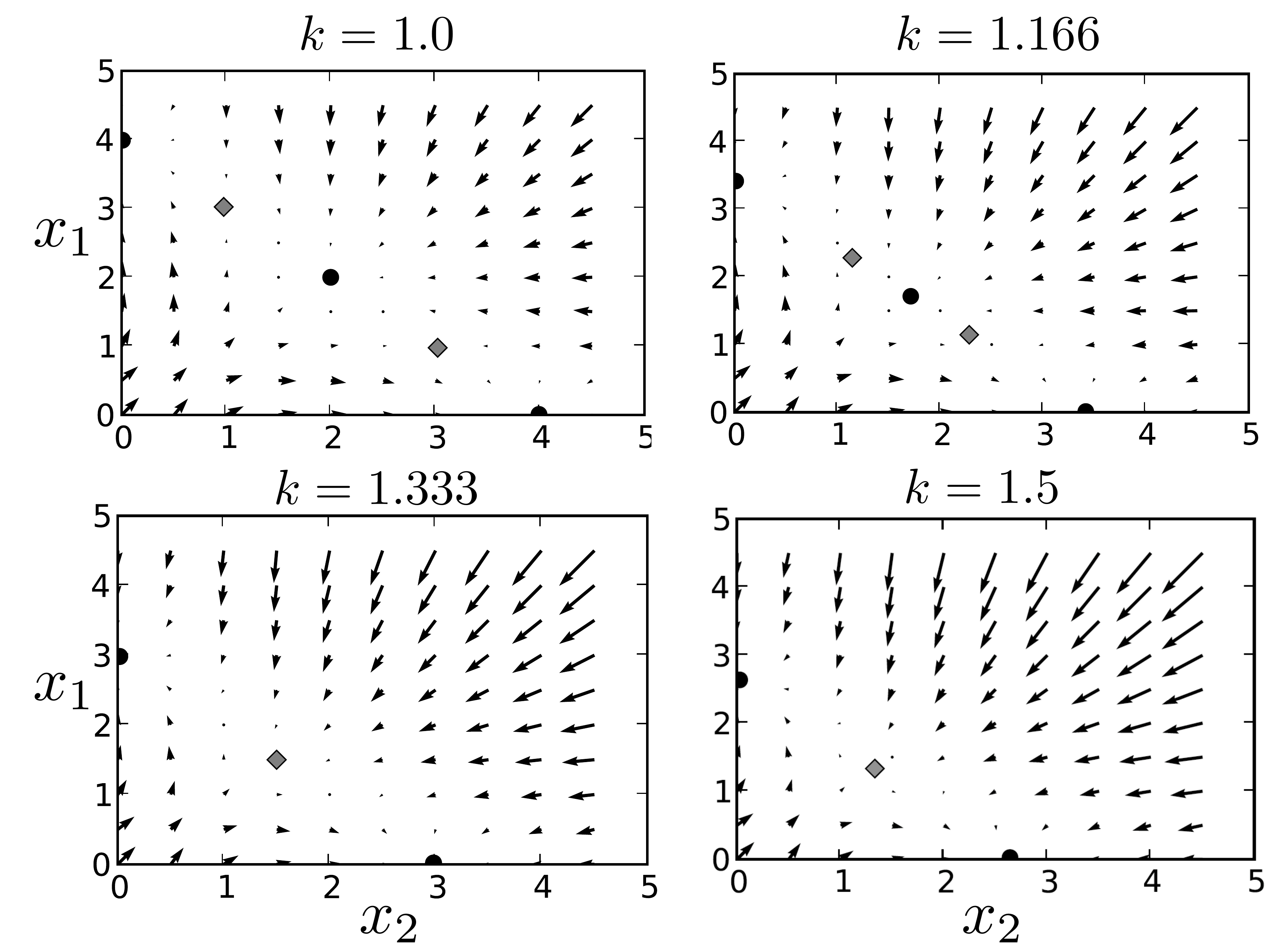}
\caption{\label{trans2} Each panel shows the state space of the two gene toggle switch for the parameter values chosen ($a = 2$ and $k = 1.0, 1.166, 1.333, 1.5$). For $k = 1.0$ there are three stable fixed points in the dynamics (shown as black dots) and two unstable fixed points (shown as gray diamonds). The arrows give the trajectory of the system at each point in phase space, with longer arrows indicating faster movement and shorter arrows slower movement. As $k$ changes from $1$ to $1.5$ the module undergoes a subcritical pitchfork bifurcation \cite{Strogatz_1994} (shown in more detail in Fig. \ref{hyster}). The two unstable fixed points converge on the interior stable fixed point, eventually destabilizing it when $k \approx 1.3$ and leaving only the two exterior stable fixed points.}
\end{figure}

For $a = 2$ and $k = 1$, the two variable system is ``tri-stable", meaning that there are three stable fixed points in the dynamcs. This state space is shown in the top left panel of Fig \ref{trans2}. The three stable fixed points are given by black dots: there is an intermediate fixed point, characterized by moderate amounts of $x_1$ and $x_2$, and two exterior fixed points, characterized by large amounts of one variable, and small amounts of the other. The gray diamonds represent the two unstable fixed points. Leaving the parameter $a = 2$ fixed and tuning $k$ from $1$ to $1.5$ drives a bifurcation from tri-stablity to bi-stabliltiy, as shown in the subsequent three panels of Fig. \ref{trans2}. The intermediate fixed point is destabilized, leaving only the two exterior stable fixed points. The bifurcation also occurs if the overall transcription rate, $a$, is increased while $k$ is kept constant---only the ratio of these parameters has a real effect on the qualitative dynamics. Their absolute values merely set the time scale.

\begin{figure}
\includegraphics[scale=.21]{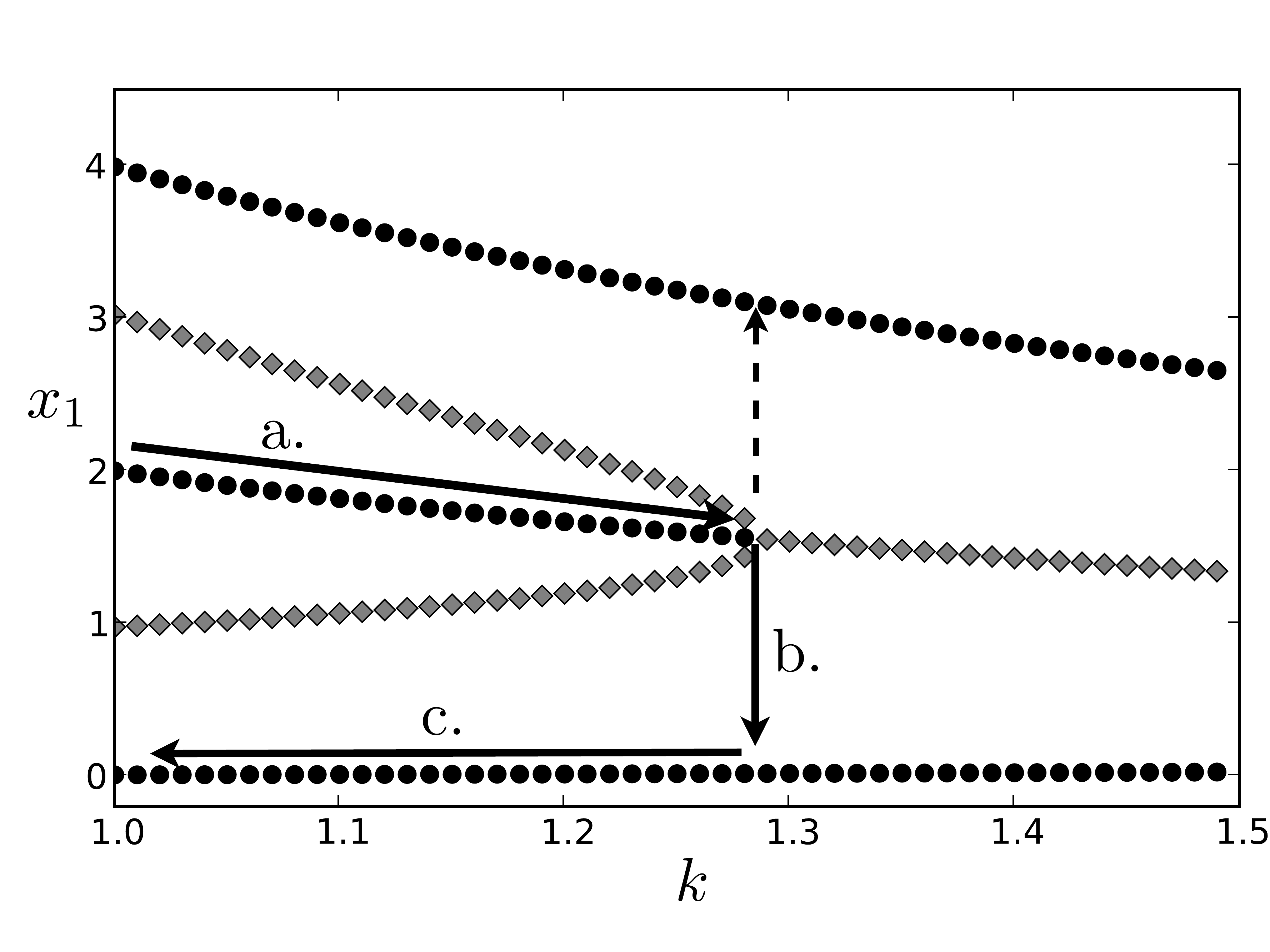}
\caption{\label{hyster} The $x_1$ position of the fixed points as the parameter $k$ changes from $1.0$ to $1.5$. The black circles represent stable fixed points while the gray diamonds represent unstable fixed points. As the module undergoes a subcritical pitchfork bifurcation, a cell at the interior stable fixed point moves along the trajectory shown by the arrow \textbf{a}. When $k \approx 1.3$ the fixed point is destabilized and the cell is pushed by any random fluctuations from the now unstable fixed point to one of the exterior fixed points following either trajectory {\bf b} or the dotted trajectory. Once the cell has been drawn to one of the exterior fixed points, if the module reverses the bifurcation and once again becomes tri-stable, the cell will remain at its new stable fixed point, following trajectory {\bf c}. So once the state of the system has undergone the bifurcation, it cannot easily be driven back to its starting position, thus making the differentiation event modeled by this bifurcation one-way.}
\end{figure}

In their analysis of the tri-stable switch, Huang {\it et al.} \cite{Huang_2007} argue that the ``interior" stable fixed point, characterized by intermediate expression levels of both genes, represents the hematopoietic stem cell, and the two ``exterior" stable fixed points represent the two committed lineages. They provide evidence that the differentiation from hematopoiesis to the erythroid or myelomonocytic lineages is caused by the destabilization of the intermediate fixed point via a subcritical pitchfork bifurcation. This in turn forces the cell to shift to one of the two other fixed points. In Fig. \ref{hyster} we show the bifurcation diagram corresponding to this event. The black dots indicate the stable fixed points, and the gray, the unstable fixed points. The $x$-axis shows the value of the bifurcation parameter, $k$. When $k = 1.0$ the circuit is tri-stable: there are three stable fixed points. When $k \approx 1.3$, the interior stable fixed point is destabilized by the approach of the two unstable fixed points. Thus, the state of the system will eventually migrate to one of the two remaining stable fixed points under perturbation. However, once the system sits at an exterior fixed points, it will not return to the interior fixed point if we reverse the process and decrease the bifurcation parameter---the exterior fixed points remain stable to small perturbations throughout. Thus, to reverse the cell differentiation, not only would the interior fixed point need to be re-stabilized; the state of the system would also need to be driven from the exterior fixed point. This ensures that cell differentiation is directional: a simple, transient, one parameter perturbation (increase in $k$) can drive the system toward specificity, but there is no similarly simple way to de-differentiate back to multipotency.

We chose the tri-stable switch proposed by Huang {\it et al} for its simplicity and empirical support. Several other models have been proposed for the genetic control circuit in hematopoiesis \cite{Bokes_2009, Cinquin_2005, Roeder_2006}. Each of these represents a reasonable model for a genetic switch, and would most likely function as desired if used as modules in the hierarchical, modular network proposed here. This underscores a key strength of the top-down approach: important properties of the dynamics are driven by the hierarchical architecture and may be insensitive to many of the specifics of each module.

\subsection{Stochastic model}

Noise is ubiquitous in biological systems and important for many functions \cite{Elowitz_2002}. The GRN as a whole, but especially the genes responsible for the process of development (and cell differentiation), must be organized for reliable performance. Thus we require that our model behaves robustly in the presence of noise. Feedback loops \cite{Eldar_2002} and modularity \cite{Variano_2004} have been shown to contribute to robustness, and because our model makes use of these design characteristics, we find that robustness is easily attained. We simulate noise in the dynamics using a Langevin equation:
\begin{equation}
  dx_i = F_{i}({\bf x}) \ dt + g \ dW_i
   \label{stochastic}
\end{equation}
Here, $F_i({\bf x})$ is the same function as in the deterministic dynamics. In the second term, $W_i$ is a Wiener process (an independent white noise process). This process introduces additive noise with a magnitude given by the parameter $g$ and with no dependence on the state ${\bf x}$.  Such noise was proposed by Kepler and Elston \cite{Kepler_2001} for models of genetic regulatory interactions. To simulate the forward evolution of the system for a length of time $\Delta{t}$, we must integrate the Wiener process over the interval which gives us $\sqrt{\Delta{t}} \ \xi_i$, where $\xi_i$ is a Gaussian random variable with zero mean and unit variance. Intuitively, the magnitude of the integral grows like $\sqrt{\Delta{t}}$ because it behaves like a one dimensional random walk: for a $1$-d random walk, the expected displacement after $n$ steps is proportional to $\sqrt{n}$. Thus we can compute the new state of the system after a time step $\Delta{t}$ using the Forward Euler integration scheme:
\begin{equation}
  x_i(t + \Delta{t}) = x_i(t) + \Delta{t} \ F_i({\bf x}) + \sqrt{\Delta{t}} \ g \ \xi_i
   \label{stochastic2}
\end{equation}
Although the Forward Euler scheme is only first order accurate, the equations are quite simple and simulating them is very fast. Thus, we can easily choose $\Delta{t}$ to be small enough to ensure accuracy. Naturally, higher order schemes could be applied as needed. Note that since the variables in our system describe the concentration of protein products, we require that no variable drop below zero: a negative concentration is non-physical. 

\begin{figure}
\includegraphics[scale=.21]{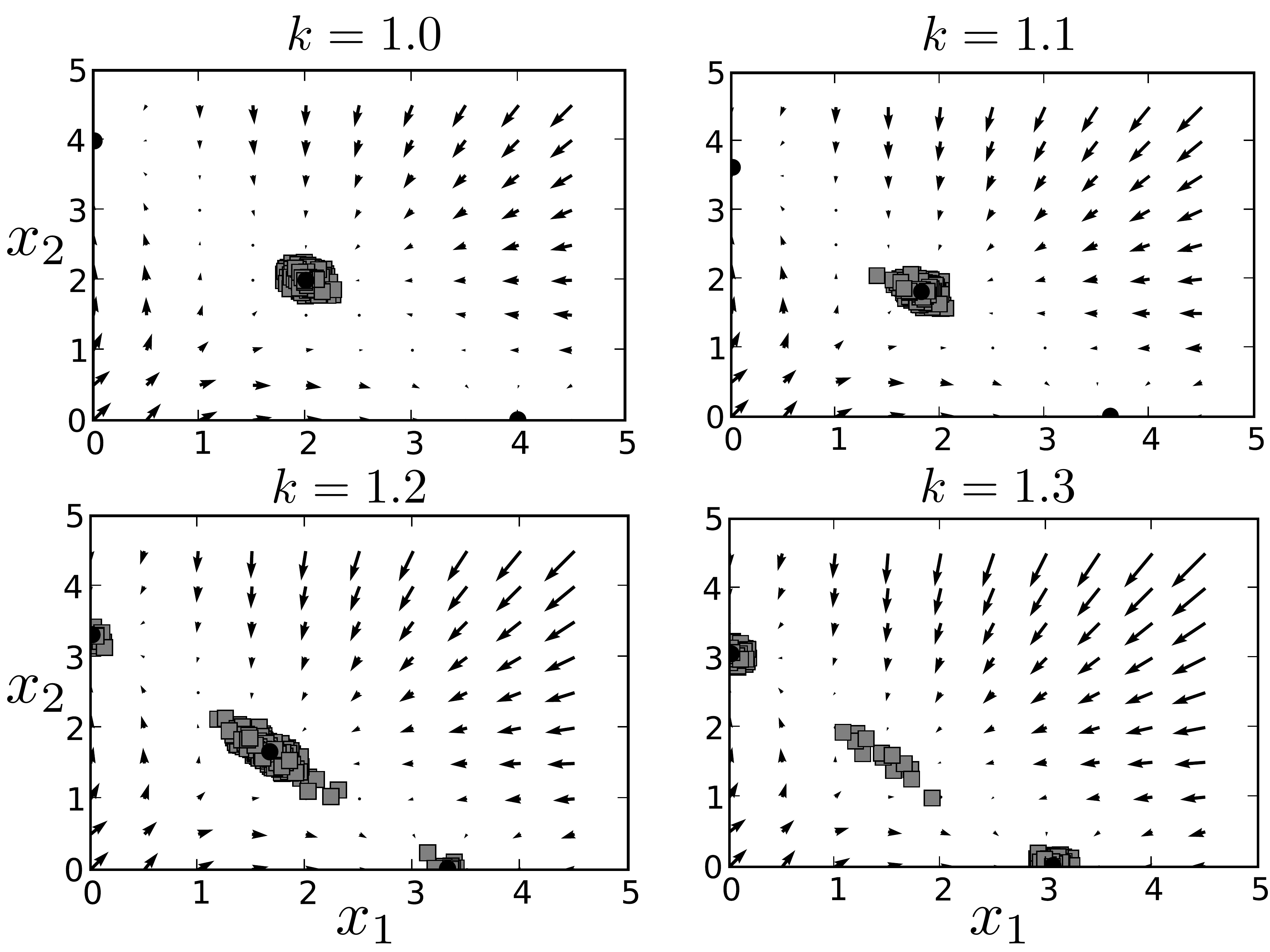}
\caption{\label{bifurcation-noise} The behavior of the module in the presence of noise as it approaches the bifurcation. Each panel shows the final positions of 500 simulations (``cells") starting from the interior stable fixed point simulated for $40$ units of $t$ (in $\Delta{t} = .001$ time steps). The noise magnitude is $g = .1$ and the overall activity $a = 2.0$ for various values of $k$.  Black circles show the positions of the stable fixed points and gray squares show the end points of each simulation.  As $k$ increases, the variance of expression levels of $x_1$ and $x_2$ increases near the interior stable fixed point. For $k = 1.0, 1.1$ none of the simulated cells ``escape" from this fixed point. For $k = 1.2$ approximately 50 cells (or $10\%$) escape, divided evenly between the two exterior fixed points. When $k = 1.3$ approximately $90\%$ escape. In similar simulations at the same noise strength, $g =0.1$, none of  the ``cells" starting near one of the exterior fixed points transitioned to the interior fixed point for any parameter values tested.}
\end{figure}

In the presence of noise, we find that the transition between stability and instability for the interior stable fixed point is no longer abrupt. As the module approaches the bifurcation point, the restoring force which pushes the state of the system back towards this interior fixed point weakens. We simulated 500 cells starting at each of the three stable fixed points for values of $k = 1.0, 1.1, 1.2, 1.3$ with $g = 0.1$ and $\Delta{t} = 0.001$ for $40,000$ time steps ($40$ units of $t$, a fairly lengthy period in simulation time relative to the timescale of transcription and translation). The distribution of the final states for cells starting at the interior stable fixed point spreads as $k$ increases (shown in Fig. \ref{bifurcation-noise}), i.e. as the value of $k$ increases, the variance of the gene expression increases. Our model hence makes a sharp prediction that may soon be testable {\it in vivo}, using flow cytometry to measure the approximate expression level of a gene for each cell in a population of cells approaching differentiation. If our model is correct, the expression level of significant genes should become more varied across the population as the cells approach a cell fate decision.  Thus, observing the variance of genes as cells differentiate would provide a clearer picture of which genes are controlling cell fate decisions: those with higher variance.  

Before the actual bifurcation point is reached, we see non-trivial numbers of cells have already moved to one of the exterior, `differentiated' fixed points. This indicates that for values of $k \gtrsim 1.2$ the interior stable fixed point becomes metastable (weakly stable). Thus the state of the system will not remain near this fixed point indefinitely in the presence of noise. As we discuss later, progenitor cell types may be metastable fixed points in which the state of the GRN rests for a short time before continuing its differentiation. This would explain the difficulty of culturing and maintaining such cell types {\it in vitro}.

In closing this section, we note that our model provides a potential dynamical explanation of the observed prevalence of binary (as opposed to trinary, etc.) branches in cell differentiation. The subcritical pitchfork bifurcation found in our modules arises from the topology of the regulatory interactions and is present for many choices of parameter values without careful tuning. This bifurcation leads naturally to binary cell differentiations.  In general, only one of the eigenvalues of the interior fixed point becomes positive at the bifurcation. This implies that flows along the other eigenvectors near the interior fixed point still push the state of the system toward that fixed point. This mechanism concentrates flows along the unstable eigenvector. In Fig. \ref{bifurcation-noise} the distribution of final states does not spread evenly through the state space; it spreads preferentially along trajectories following the two sides of the unstable manifold of the interior fixed point. Near the fixed point, the unstable manifold is tangent to the unstable eigenvector; farther from the fixed point, the unstable manifold need not be a straight line. These trajectories, in turn, lead to one of the two exterior fixed points. Trinary branchings, since they cannot depend on such simple properties of the state space, are comparatively much more fragile and require correspondingly careful selection of parameters. We have found in our model that binary branching occurs `for free' from a robustly occurring pitchfork bifurcation, with the simulated cells flowing away from their starting state along one of the two ends of the single, destabilized eigenvector.

\section{Sequential Differentiation}

\begin{figure}
\includegraphics[scale=.21]{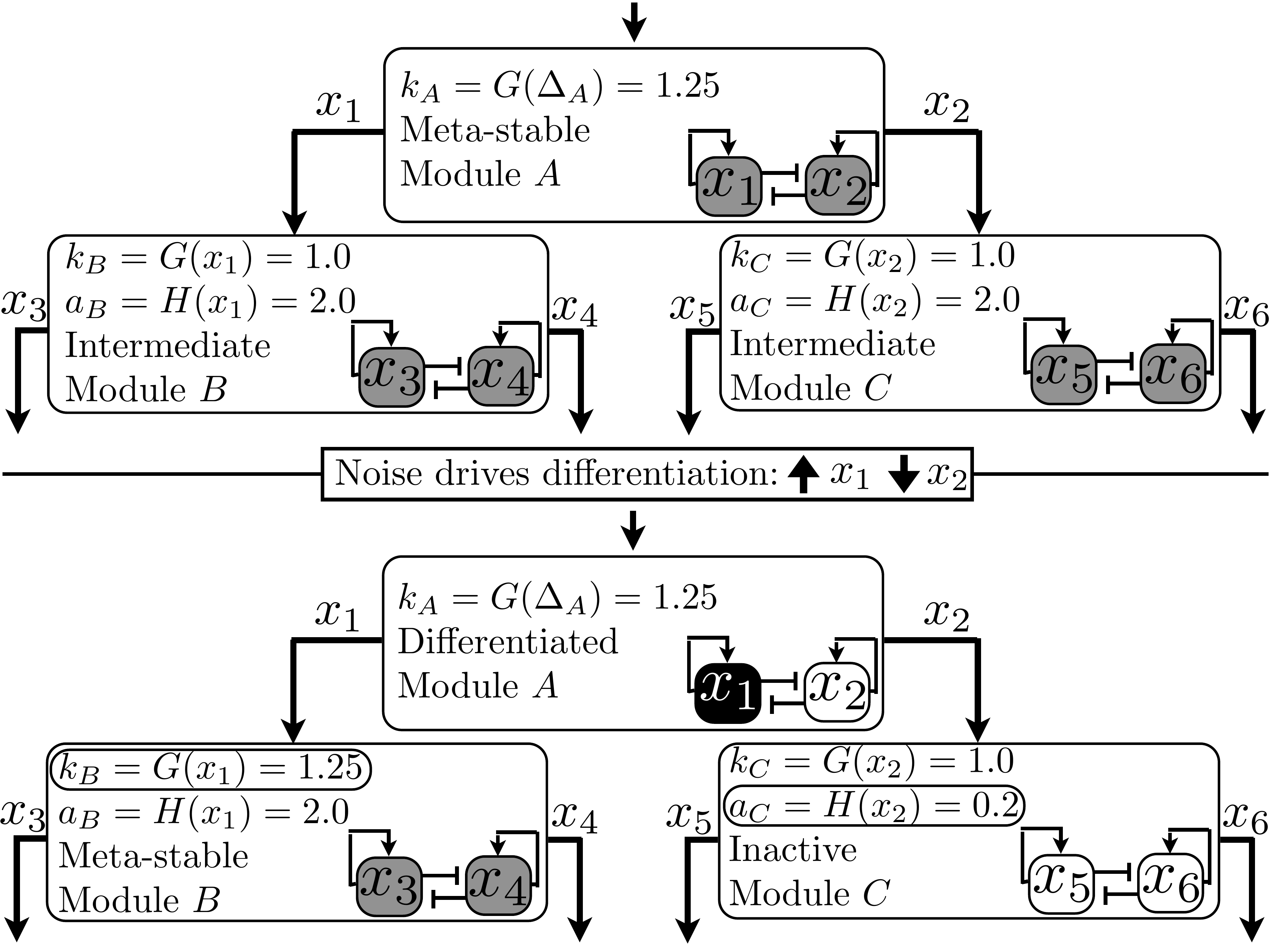}
\caption{\label{celldiff} The modular network and consequences of a single cell fate decision, in detail. The $k$ and $a$ parameters in all three modules are functionally dependent on one of the genes in the module directly above them. In the case of module A, we denote the stimulus as a generic external signal $\Delta_A$ to reflect the fact that it could come from another gene, or some external differentiation signal. In the top panel, the state of the system is poised in the intermediate, meta-stable state of module A ($k_A = 1.25$). Modules B and C are both in the intermediate stable state, expressing $x_3$, $x_4$, $x_5$, and $x_6$ at a moderate level. In the lower panel, module A has undergone differentiation, driven by noise, switching to one of the two exterior stable states, $x_1$ high, $x_2$ low. This in turn has caused module B to become meta-stable by increasing the value of $k_B$ from $1.0$ to $1.25$. Module C is inactivated by the change in parameter $a_C$ from $2.0$ to $0.2$, essentially silencing the genes in the module.}
\end{figure}

\begin{figure}
\includegraphics[scale=.21]{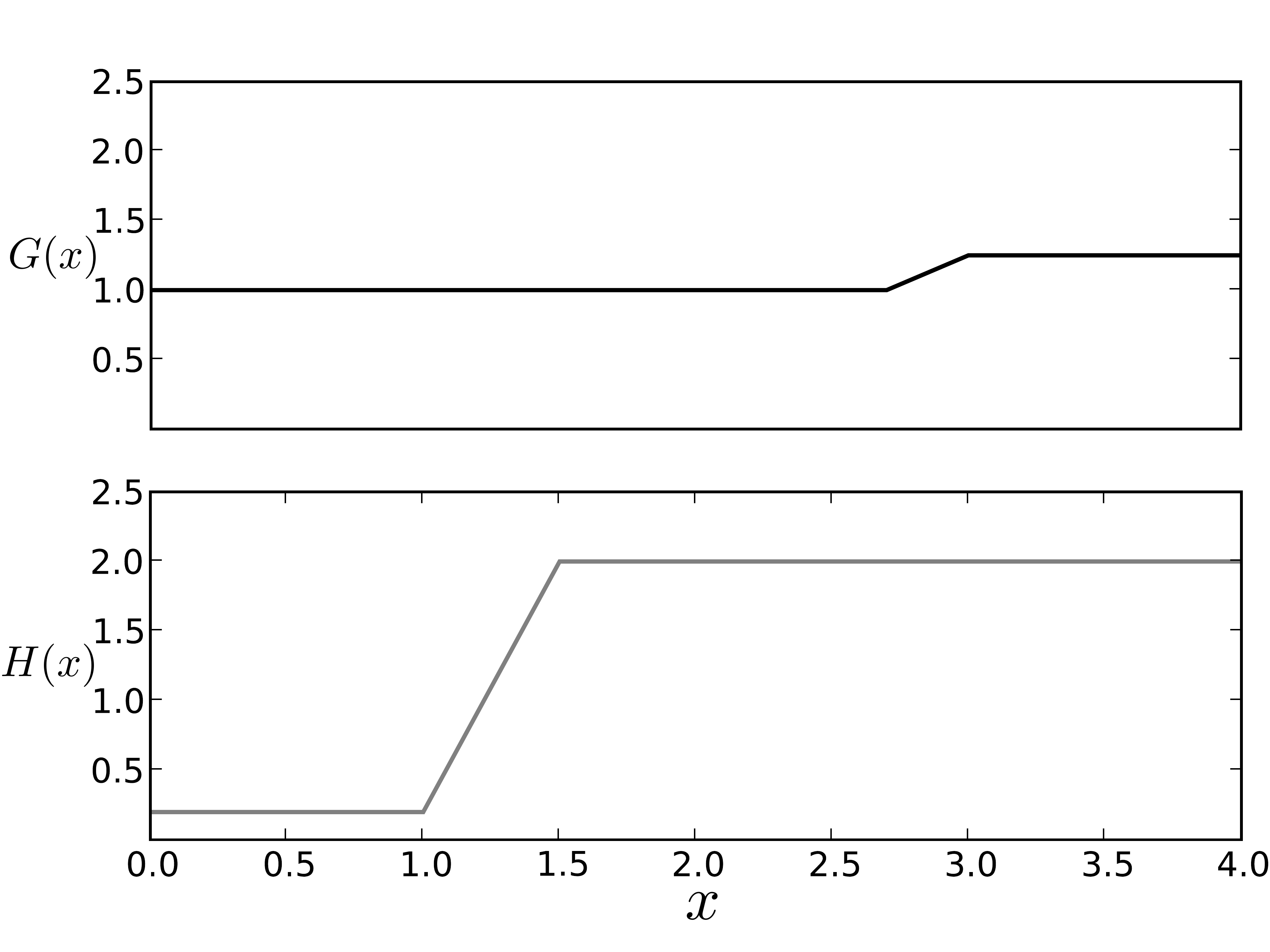}
\caption{\label{GandH} The piecewise linear response functions $G$ and $H$. In the top panel we show the function $G$, which controls the decay rate $k$, and in the bottom panel we show the function $H$, which controls the transcription rate $a$. Specifically, $G = 1.0$ for $x \leq 2.7$ and $G = 1.25$ for $x \geq 3.0$, interpolating between these values for $2.7 \leq x \leq 3.0$. Similarly $H = 0.2$ for $x \leq 1.0$ and $H = 2.0$ for $x \geq 1.5$, interpolating between these values for $1.0 \leq x \leq 1.5$.}
\end{figure}

Now we are ready to build the full hierarchical model. In Fig. \ref{celldiff}, we show a differentiation event in more detail: three modules are shown, $A, B,$ and $C$. Module $A$ influences $B$ and $C$. In the top panel, module $A$ is in the intermediate fixed point, but this fixed point is only meta-stable. This meta-stability is caused by the influence of the signal $\Delta_A$. If module $A$ is part of a larger cascade, then the signal will originate from a gene in the module above $A$ in the hierarchy; if $A$ is at the top of the hierarchy then the signal may originate from outside the cell. Module $A$ expresses moderate levels of $x_1$ and $x_2$, which in turn influence the parameters $k_B, a_B$ and $k_C, a_C$ in modules $B$ and $C$, through the functional relationships:
\begin{equation}
\begin{array}{l}
\displaystyle k_B = G(x_1), k_C = G(x_2) \\
\displaystyle a_B = H(x_1), a_C = H(x_2)\\
\end{array} 
\label{functions}
\end{equation}
For simplicity, the same functions $G$ and $H$ are used throughout the network. In Fig. \ref{GandH} we show how the piecewise linear functions $G$ and $H$ respond to changes in the variable on which they depend. The specific functional form chosen is not important; sequential differentiation will occur as long as the functions satisfy two basic requirements. First, if the value of $x$ increases substantially, then the interior fixed point in the influenced module should be destabilized, preparing it to undergo differentiation. Thus, the function $G(x)$, which controls $k$, should increase slightly in value in response to higher values of $x$. This higher value of $G(x)$ will drive the bifurcation above. Second, if the value of $x$ decreases dramatically, indicating that the gene $x$ has been shut off, then the transcription rate, $a$, of the downstream module, which is under the influence of $x$, should be dramatically decreased. To achieve this, for small values of $x$, $H(x)$ should be small, effectively silencing the genes in the influenced module.

Returning to Fig. \ref{celldiff}, moderate levels of $x_1$ and $x_2$ result in parameters in modules $B$ and $C$ which maintain stable, moderate expression levels of their genes $x_3$, $x_4$, $x_5$ and $x_6$. Note that any modules downstream of $B$ and $C$ would also express their genes at an intermediate level. Since module $A$ is in a meta-stable state, the state of the system is driven by noise after some time to one of the two exterior fixed points; in the Figure, $x_1$ high, $x_2$ low. The result is shown in the lower panel: module $C$ has been silenced by the decrease in expression of $x_2$ (thus also shutting off all the genes in modules downstream from it, under the control of $x_5$ or $x_6$). Module $B$, under the influence of increased $x_1$, has become meta-stable, and is thus waiting to drive a differentiation event as the system ``chooses" between expressing $x_3$ and $x_4$.

The proposed links integrating the modules correspond to the simplest possible regulatory connections that generate the observed branching behavior. In most models, regulatory control acts on the process of transcription. However, the decay rate $k$ can also act as a control parameter, if one protein promotes the degradation of another protein. In the preceding discussion, we chose to use both $k$ and $a$ as control parameters to simplify the model. Since the qualitative behavior of the modules depends only on the relative values of the decay $k$ and the activation $a$, the desired behavior can be accomplished using only the activation rate as a parameter. This requires that the activation rate in a regulated module respond non-monotonically to the relevant gene in the controlling module: for increased levels of the relevant gene, the activation rate decreases slightly (which has the same effect as increasing $k$), whereas for decreased levels of the relevant gene, the activation rate drops dramatically, as it does for $H(x)$, to silence the module. Precisely what regulatory connections appear in real networks remains to be seen; however, modules of mutually regulating transcription factors clearly are linked in real GRNs.

Thus our hierarchical network model allows a single, multipotent cell type to access a multitude of distinct, terminal states in a biologically sensible way. Cinquin and Demongeot \cite{Cinquin_2005} present an alternative mathematical model: a single high-dimensional genetic switch with more than two possible outcomes. Their model produces promiscuous expression of antagonistic factors (which represents the undifferentiated state), whereupon a symmetry breaking bifurcation results in increased expression of one of the genes in the circuit, and decreased expression of all the others. While Cinquin and Demongeot argue that such multi-switches are necessary to capture particular biologically observed differentiations, as a generic mechanism for generating a diversity of terminal cell types, the modular, hierarchical approach taken in our model is more robust: it is not reliant on the highly symmetric, all-to-all coupling present in the Cinquin model. Our model should work with a diversity of different modules, so long as they conform to certain, loose criteria. Although we imposed an $x_1$, $x_2$ symmetry on the Huang circuits used in this paper, we did so only to reduce the number of parameters and simplify the discussion: the Huang switch maintains tristability, and all its other important properties, when the parameters are asymmetric \cite{Huang_2007}.

There has been much interesting mathematical work on symmetry breaking in biological systems \cite{Dias_2006, Elmhirst_2004, Stewart_2003}.  ``Coupled cell systems" \cite{Dias_2006} present a potentially interesting mathematical toolkit for analyzing the dynamics of modular networks, such as the network proposed in this paper: dynamical components called ``cells" interact with one another according to a given network of connections. Studying the hierarchical network architecture proposed in this paper using the tools of coupled cell systems may provide useful dynamical insights, and is a fruitful avenue for future work.

\section{Whole network results}

\begin{figure}
\includegraphics[scale=.21]{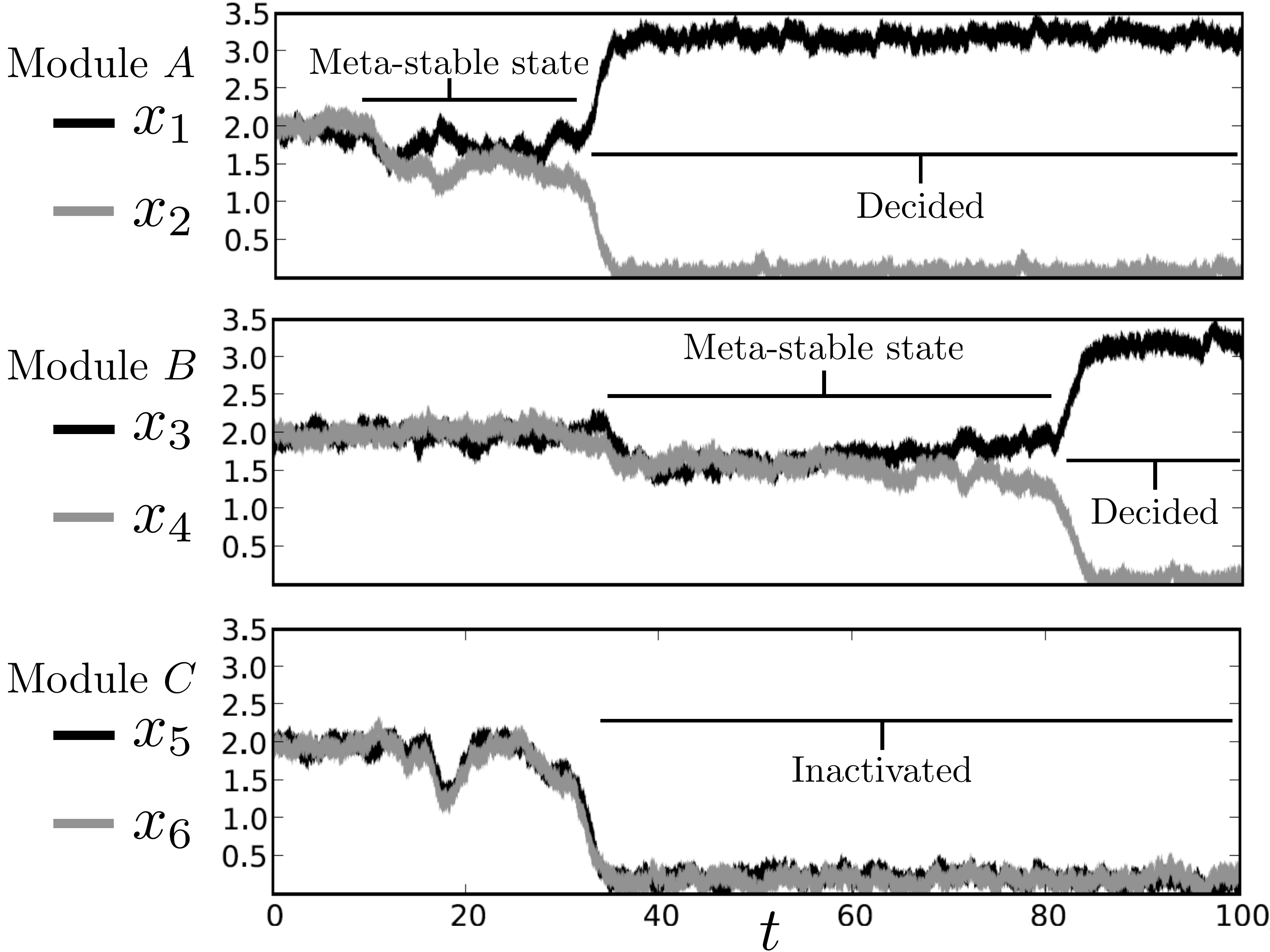}
\caption{\label{StochRes} A simulation of three interacting modules with noise $g = 0.1$. At $t = 5$ the interior fixed point of module $A$ is changed from stable to meta-stable by an external stimulus. For the next $\approx 10$ units of time $t$ the top module remains meta-stable, then finally falls into the $x_1$ high exterior fixed point, as shown in the top panel. This silences module $C$ (and thus genes $x_5$ and $x_6$, which are rendered inactive) as shown in the bottom panel. Finally, in the middle panel we see module $B$, which became meta-stable after module $A$ made its cell fate decision and entered the exterior fixed point. Module $B$ remains meta-stable for $\approx 25$ units of $t$ until it transitions to an exterior fixed point, completing the differentiation of the cell through this segment of the network.}
\end{figure}

Fig. \ref{StochRes} shows the result of a simple three module simulation, with the noise parameter $g = 0.1$. The differentiation tree functions properly in the presence of noise: at the outset, all genes are expressed at a moderate level. At time $t = 5$, a signal is introduced which destabilizes the top module. After some time, the state of the the top module is decided in favor of $x_1$. As $x_2$ decreases, we see that genes $x_5$ and $x_6$ are also silenced. The genes $x_3$ and $x_4$ continue to be expressed at moderate levels for some time, until through random fluctuations, the system exits the meta-stable state, resulting in increased $x_4$ expression and decreased $x_3$ expression. 
The individual cell fate decisions are strongly stochastic in the model as presented. Once the interior fixed point is destabilized, random fluctuations determine into which exterior fixed point the state of the system falls.  This is consistent with the stochastic component of cell fate decisions observed in previous work \cite{Graf_2002, Hume_2000}. As noted in Huang {\it et al.} \cite{Huang_2007}, the perturbation from the interior fixed point can be done asymmetrically, resulting in ``intrinsic" as opposed to stochastic cell fate decisions.

At the outset, when any outcome was possible and the cell multipotent, all of the genes are expressed. As the cell differentiates and loses its pluripotency, the genes which correspond to the lineages which are no longer accessible to the cell are silenced. This sequence of silencing events provides a clear explanation of the exclusivity of cell differentiation, as well as a mechanism for ensuring an orderly progression of cell types.  Consider a differentiation tree with $n$ levels and assume the tree is complete and symmetric, so that $n-1$ binary decisions occur. Fig. \ref{Biftree} is an example of such a tree for $n=2$. A general tree will contain $(2^n -1)$ modules and $N = 2(2^n -1)$ genes.  At the beginning of the differentiation process all $N$ genes will be expressed promiscuously at intermediate levels. When the module at the top of the tree transitions to one of its two possible exterior fixed points, the cascade of silencing turns off an entire $(n-1)$-level sub-tree.  Thus the effective dimension of the state space is reduced from $N$ to $(N - 2(2^{n-1}-1)-1)$. The silenced genes are fixed near $0$ and subsequent time evolution of the system can only take place in the remaining state space, which is associated exclusively with the available lineages.  Exclusivity and orderly cell type progression follow immediately from this successive restriction of the dynamics to smaller and smaller subspaces.  This picture also yields an experimental prediction.  If the dynamical mechanism proposed here is correct, then differentiation from multipotent to more specific cell types should be characterized by a sequence of gene silencing events, with many more genes silenced (on average) early in the differentiation process and fewer genes silenced as the terminal type is achieved.

As seen in the simulation described above, individual cells can remain in the meta-stable state for a long time before continuing their differentiation. We simulated $2000$ iterations of the model, and measured the length of time the system spent in meta-stable states. We found an exponential distribution of waiting times, with some of our simulated cells spending as many as $290$ units of $t$ in the meta-stable state. For the purposes of comparison, we note that once the state escapes the region of state space close to the meta-stable fixed point and the transition to an exterior fixed points begins, the transition often occurs in less than $5$ units of $t$. So while individual cells cannot be maintained in these meta-stable states indefinitely, such states can be quite long-lasting. We have not considered the proliferation of cells (i.e. division maintaining cell type), as described by Hoffmann {\it et al} \cite{Hoffmann_2008}. The proliferation rate is a function of the state of the GRN. Thus if the proliferation rate is high in the meta-stable states of intermediate gene expression, a population of cells in the meta-stable state could be maintained indefinitely. The number of progenitor cells found {\it in vivo} is the result of a balance between: cells differentiating into the meta-stable progenitor state from more multipotent cell types; progenitor cells proliferating; and progenitor cells differentiating out of the progenitor cell type to a more lineage-committed state through the stochastic dynamics described above. Thus the meta-stability of the dynamical fixed point may provide an explanation for the limited capacity for self-renewal of some progenitor cell types {\it in vitro}.

\section{Conclusion}
Our model reproduces all of the desired behaviors of cell differentiation by using established network properties of real genetic networks. Cell differentiation occurs between dynamically distinct, stable cell types. Cell fate decisions are directional: small parameter changes can easily drive differentiation down the tree, but not back up again.  Differentiation occurs in a branching, exclusive fashion, due to the properties of the bifurcation and the progressive restriction of available state space. And finally, a cell type higher in the differentiation tree exhibits promiscuous expression of genes specific to all its accessible lineages, by construction. It is easy to add additional control mechanisms to this model; for example, we could require external signals, in concert with the influence of ``higher" modules, to drive each bifurcation. Our goal, however, is merely to outline a minimal dynamical system capturing certain generic features of cell differentiation.

In addition, our model suggests a new role for modularity and an explanation for the prevalence of binary cell fate decisions. Modules allow for the easy implementation of repeated behaviors, in this case facilitating the repeated branchings observed in cell fate decisions. The regulatory structure of the hierarchical, modular network topology serves to channel trajectories in the state space, by progressively restricting them to lower dimensional subspaces. This guides the state of the system through an orderly progression of fixed points, representing the various cell types. Further, binary cell fate decisions emerge effortlessly and robustly from the topology of our network modules due to the behavior of flows near the interior fixed point of intermediate gene expression. The dynamical simplicity and robustness of binary cell fate decisions could explain their prevalence in cell differentiation. Our model thus illustrates that many of the important properties of cell differentiation can be captured quite naturally by a system of loosely coupled genetic modules.

Building progressively more accurate whole-system models that recapitulate key functionalities provides an important complement to the traditional ``bottom up" approach, which seeks maximal integration of all available details. A ``top down" approach, focused on less detailed minimal models, allows us to focus on understanding the influence of the topology of the genetic regulatory network on the resulting dynamics. The details of our individual modules and their interconnections are of course oversimplified and incomplete. However, emerging evidence from the topology of real genetic networks suggests that the overall modular architecture is likely to be correct. We hope that future experimental results will show to what degree the genetic regulatory network in general, and the cell differentiation network in particular, can be treated as a collection of loosely coupled modules.  Our model makes two clear predictions regarding the cell differentiation network. On the one hand, the variance in expression level of genes involved with the differentiation process should increase across a population of cells as the population approaches differentiation. On the other, the progressive restriction of available state space should be reflected in a sequence of gene silencing events as cells proceed from more multipotent, promiscuously expressing types down a particular lineage, with fewer genes (on average) going silent as the terminal cell type is approached. As large-scale models improve through interaction with experimental evidence, we believe that the top down and bottom up approaches can eventually ``meet in the middle", fitting carefully specified subsections of the genetic network into a larger network framework.

\end{document}